\documentclass[10pt,conference,twoside]{IEEEtran}
\pdfoutput=1
\usepackage{cite}
\usepackage{graphicx}
\usepackage{psfrag}
\usepackage{amsthm}
\usepackage{url}
\usepackage{amsmath}
\usepackage{array}
\usepackage{amssymb}
\usepackage{authblk}
\usepackage{amsfonts}
\usepackage{graphicx}
\usepackage{epstopdf}
\usepackage{algorithm}

\newtheorem{lemma}{Lemma}

\newtheorem{corollary}{Corollary}

\newtheorem{theorem}{Theorem}

\newtheorem{example}{Example}

\usepackage{setspace}
\usepackage{amscd}
\usepackage{amsthm}
\usepackage{mathrsfs}
\usepackage{epsfig}
\usepackage{color}
\usepackage{textcomp}
\usepackage{multirow}
\usepackage{caption}
\usepackage{threeparttable}
\title{On the Number of Optimal Index Codes}

\begin{document}
\author{Kavitha. R and B. Sundar Rajan
}\affil{Dept. of ECE, IISc, Bangalore 560012, India, Email: kavithar1991@gmail.com, bsrajan@ece.iisc.ernet.in.}
\date{\today}
\renewcommand\Authands{ and }
\maketitle
\thispagestyle{empty}	
\begin{abstract}
In Index coding there is a single sender with multiple messages and multiple receivers each wanting a different set of messages and knowing a different set of messages a priori. The Index Coding problem is to identify the minimum number of transmissions (optimal length) to be made so that all receivers can decode their wanted messages using the transmitted symbols and their respective prior information and also the codes with optimal length. Recently in \cite{Ourwork}, it is shown that different optimal length codes perform differently in a wireless channel. Towards identifying the best optimal length index code one needs to know the number of  optimal length index codes. In this paper we present results on the number of optimal length index codes making  use of the representation of an index coding problem by an equivalent network code.  We give the minimum number of codes possible with the optimal length. This is done using a simpler  algebraic formulation of the problem compared  to the approach of Koetter and Medard \cite{KoMe}. %We also give a method to identify the best linear solution in terms of minimum-maximum error probability among all codes with the optimal length.
\\
%Single unicast problem is the special case where each receiver wants a single unique messages.
\end{abstract}

%\begin{IEEEkeywords} 
% Index Coding, Network Coding.  
%\end{IEEEkeywords}

\section{Introduction} 
\IEEEPARstart{W}{e} consider the index coding problem first introduced by Birk \textit{et. al.} in \cite{Birk}. In an index coding (IC) problem, there is a single sender with multiple messages and some receivers. Each of them wants a set of messages and knows a set of messages a priori. A single uniprior IC problem is a scenario where each receiver knows a single unique message a priori and a unicast problem is another where each receiver wants a unique set of messages. A single unicast is when the size of each of those wanted sets in a unicast problem is one. One needs to identify the minimum number of transmissions to be made so that all receivers can decode their wanted messages using the transmitted bits and their respective prior information. Ong and Ho in \cite{OMIC} proposed the optimal length of a uniprior index coding problem. El Rouayheb \textit{et. al.} in \cite{PoYo1} found that every index coding problem can be reduced to an equivalent network coding problem. An algebraic representation of network codes was done by Koetter and Medard in \cite{KoMe}  . In this paper we present an algebraic characterisation of an index code after reducing it to an equivalent network code. Harvey \textit{et.al} in \cite{matrixcom} proposed an algorithm for network codes for multicast problems, which is based on a new algorithm for maximum-rank completion of mixed matrices. Our problem is not a multicast problem. Hence the results in \cite{matrixcom} cannot be applied.

There can be several linear optimal index codes  in terms of lowest number of transmissions for an IC problem. But among them one needs to identify the index code which minimizes the maximum number of transmissions that is required by any receiver in decoding its desired message \cite{Ourwork}. The motivation for this is that each of the transmitted symbols is error prone in a wireless scenario and lesser the number of transmissions used in decoding the desired message, lesser will be its probability of error. Hence among all the codes with the same length, the one for which the maximum number of transmissions used by any receiver is the minimum, will have minimum-maximum error probability. This has already been discussed in \cite{Ourwork} where the solution for uniprior case is found.  %We give a method to find the best linear solution in terms of minimum-maximum error probability among all codes with the optimal length for a single unicast case in Section III-C.

The contributions and organisation of this paper may be summarized as follows:
\begin{itemize}
\item The paper through an algebraic characterization, gives a method to identify the optimal length of a linear solution for a single unicast index coding problem. This is done by finding a transfer matrix (whose elements depend on the index code we choose) which relates the input messages and the decoded messages. This is done in Section III. 
\item We give the minimum number of codes possible with the optimal length for a single unicast index coding problem. This is done in Section III-B. We find this by finding the minimum number of feasible solutions of a linear system of equations which represents our index coding problem.
\end{itemize}

The proofs of all the lemmas and theorems are given in Appendix along with illustrative examples.

%\textbf{\textit{Notations:}} For a matrix ${A}, {A}^T$ and ${A}^{H}$ denotes transpose and Hermitian transpose of the matrix ${A}$ respectively. 
\section{problem formulation}
\label{Problem formulation}
A general index coding problem can be formulated as follows: There are $n$ messages, $ x_{1}, x_{2}, \ldots, x_{n} $ and $m$ receivers. Each receiver wants a set of messages, $ W_{i} $ and knows a set of messages $ K_{i} $. For a  general unicast problem, $ W_{i} \cap  W_{j} =\emptyset$, for $ i \neq j $. The special case when $ m = n $ and $W_{i} = \lbrace x_{i}\rbrace $ is called a single unicast problem. A general unicast problem can always be reduced to a single unicast problem with $\mid W_{i}\mid=1$ by replication of receivers. Hence the observations in this paper applies to a general unicast problem as well. The optimal length of a linear solution of an IC problem is identified. Also, a lower bound on the total number of linear index coding solutions with the optimal length for a single unicast problem is identified. Any single unicast problem can be represented by an equivalent network coding problem as in Fig. 1. This was proposed by El Rouayheb \textit{et. al.} in \cite{PoYo1}.\\

\begin{figure}[htbp]
\centering
\includegraphics[scale=.75]{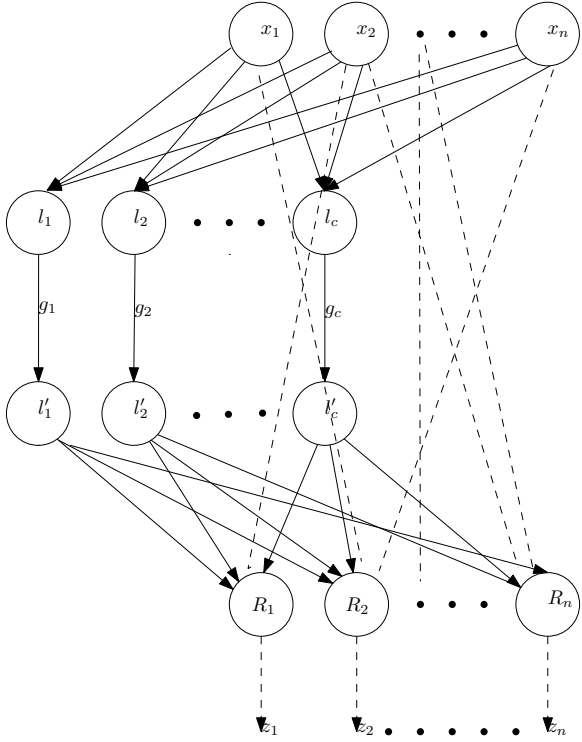}
\caption{Representation of a unicast IC problem by an equivalent network code.}
\end{figure}

\begin{figure*}
%{\footnotesize
\begin{eqnarray}
\b{Y}^T=[ Y((x_{1},l_{1}))~ Y((x_{1},l_{2}))~ \ldots ~ Y((x_{1},l_{c}))\notag\\  Y((x_{2},l_{1}))~ Y((x_{2},l_{2})) ~ \ldots ~ Y((x_{2},l_{c}))\notag\\ \vdots \notag\\  Y((x_{n},l_{1}))~ Y((x_{n},l_{2}))~\ldots ~  Y((x_{n},l_{c}))\notag\\  Y((x_{K_{1,1}},R_{1}))~Y((x_{K_{1,2}},R_{1}))\ldots~ Y((x_{K_{1,\mid K_{1}\mid}},R_{1}))\notag\\ Y((x_{K_{2,1}},R_{2}))~Y((x_{K_{2,2}},R_{2}))~\ldots ~Y((x_{K_{2,\mid K_{2}\mid}},R_{2}))\notag \\  \vdots \notag \\ Y((x_{K_{n,1}},R_{n}))~Y((x_{K_{n,2}},R_{n}))~Y((x_{K_{n,\mid K_{n}\mid}},R_{n}))] 
\label{eqn:YY}
\end{eqnarray}
%}
\hrule
\end{figure*}
Here each of the messages $ x_{1}, x_{2},\ldots, x_{n} $ is represented by a source node and $ g_{1}, g_{2},\ldots, g_{c} $ represent the  broadcast channel and $l_{1},l_{2},\ldots,l_{c},l'_{1},l'_{2},\ldots,l'_{c}$ represent the intermediate  nodes. When two or more edges have the same tail node, they carry the same message. Also $l'_{i}$ transmits to its outgoing edges whatever it gets by $g_{i}$. The source nodes transmit their respective messages as such through their outgoing edges. The length of the index code is represented by $c$. The optimal  value of $c$ among all linear solutions of an IC problem is to be found. Our operations are over the finite field $F_{2}$. But the results in this paper can be carried over to other fields also. The dashed lines represent the connection between a receiver node and its prior message (node) among the set of messages (nodes) i.e, they represent the side information possessed by the receivers. For every single unicast problem, we can find a graph like given in Fig. 1. Let us call it $G$. The graph $G$ can be represented as $G=(V, E)$, where $V=\lbrace x_{1}, x_{2}, \ldots, x_{n}, l_{1}, l_{2}, \ldots, l_{c}, l'_{1}, l'_{2}, \ldots, l'_{c}, R_{1}, R_{2}, \ldots, R_{n} \rbrace$ is the vertex set and $E$ is the edge set. We can observe that $\mid E \mid$ = $(2n+1)c + \overset{n}{\underset{i=1}{\sum}}\mid K_{i} \mid $. An edge connecting vertex $ v_{1}$ to $v_{2}$ is denoted by ($v_{1},v_{2}$) where $v_{1}$ is the tail of the edge and $v_{2}$ is the head of the edge. For an edge $e$, $Y(e)$ represents the message passed in that edge. We can get a transfer matrix $ M_{n \times n} $ (which is shown in section \ref{Algebraic Formulation}) such that $ \b{Z} = [   z_{1} ~z_{2}~ \ldots ~z_{n}  ]^T, $ the vector of output messages at each of the receivers, can be expressed as 
     \begin{equation}
     \b{Z}=  M~\b{X},
     \end{equation}
      where $\b{X}$ = $[ x_{1} ~  x_{2} ~\ldots~  x_{n}]^T$, the vector of input messages. Hence, we can solve the IC in $c$ number of transmissions if $M$ is an identity matrix.   
\section{Algebraic Formulation}
\label{Algebraic Formulation}
For a general single unicast problem, we can find a matrix $M_{n\times n}$ such that the vector of output bits $\b{Z}= M~\b{X}$. We can observe that $M$ is a product of three matrices as given in (\ref{eqn:M}).\footnote{We are not following Koetter and Medard's approach \cite{KoMe}. If we had followed their approach in a strict sense we would have got matrix $A$ of order ($\mid E \mid \times n$), $F$ of order ($\mid E \mid \times \mid E \mid$)and $B$ of order ($n \times\mid E \mid $). We give a simpler formulation for the matrices $A$, $F$ and $B$  for a given index coding problem.} We will give the structure of each of these matrices first and then explain how we derived (\ref{eqn:M}). 
\begin{equation}
M=B~F~ A
\label{eqn:M}
\end{equation}
The matrix $A$ relates the input messages and the messages flowing through the outgoing edges of all the source nodes. $A$ satisfies the following relation.\\
\begin{equation}
 \b{Y}=  A ~ \b{X},\\
 \label{eqn:Y}
\end{equation}

\begin{figure*}
{\footnotesize
 \begin{eqnarray}
 \b{Y}'^T= [Y((l'_{1},R_{1}))~Y((l'_{1},R_{2}))~ \ldots Y((l'_{1},R_{n}))\notag\\Y((l'_{2},R_{1}))~Y((l'_{2},R_{2}))~\ldots Y((l'_{2},R_{n}))\notag\\\vdots \notag\\ Y((l'_{c},R_{1})) ~Y((l'_{c},R_{2}))~\ldots ~Y((l'_{c},R_{n})) \notag \\Y((x_{K_{1,1}},R_{1}))~Y((x_{K_{1,2}},R_{1}))\ldots Y((x_{K_{1,\mid K_{1}\mid}},R_{1}))\notag\\ ~Y((x_{K_{2,1}},R_{2}))~Y((x_{K_{2,2}},R_{2})) ~\ldots Y((x_{K_{2,\mid K_{2}\mid}},R_{2})) \notag \\ \vdots \notag \\ Y((x_{K_{n,1}},R_{n}))~Y((x_{K_{n,2}},R_{n}))\ldots Y((x_{K_{n,\mid K_{n}\mid}},R_{n})) ]
  \label{eqn:YYY}
  \end{eqnarray}
}
  \vspace{-.3 cm}
\end{figure*}

  \begin{figure*}
\hrule
\vspace{.4 cm}
\begin{equation} 
\begin{tiny}
F_{B}=\left[ \begin{array}{cccccccccccccc}
\beta_{(x_{1},l_{1})}&0& \ldots &0&\beta_{(x_{2},l_{1})}&0& \ldots &0&\ldots &\beta_{(x_{n},l_{1})}&0& \ldots &0 \\
\beta_{(x_{1},l_{1})}&0& \ldots &0&\beta_{(x_{2},l_{1})}&0& \ldots &0&\ldots &\beta_{(x_{n},l_{1})}&0& \ldots &0 \\
.\\
.\\
\beta_{(x_{1},l_{1})}&0& \ldots &0&\beta_{(x_{2},l_{1})}&0& \ldots &0&\ldots &\beta_{(x_{n},l_{1})}&0& \ldots &0 \\
0&\beta_{(x_{1},l_{2})}& \ldots &0&0&\beta_{(x_{2},l_{2})}& \ldots &0&\ldots &0&\beta_{(x_{n},l_{2})}& \ldots &0\\
0&\beta_{(x_{1},l_{2})}& \ldots &0&0&\beta_{(x_{2},l_{2})}& \ldots &0&\ldots &0&\beta_{(x_{n},l_{2})}& \ldots &0\\
.\\
.\\

0&\beta_{(x_{1},l_{2})}& \ldots &0&0&\beta_{(x_{2},l_{2})}& \ldots &0&\ldots &0&\beta_{(x_{n},l_{2})}& \ldots &0\\
.\\
.\\
.\\
0&0& \ldots &\beta_{(x_{1},l_{c})}&0&0& \ldots &\beta_{(x_{2},l_{c})}&\ldots &0&0& \ldots&\beta_{(x_{n},l_{c})}\\
0&0& \ldots &\beta_{(x_{1},l_{c})}&0&0& \ldots &\beta_{(x_{2},l_{c})}&\ldots &0&0& \ldots &\beta_{(x_{n},l_{c})}\\
.\\
.\\
0&0& \ldots &\beta_{(x_{1},l_{c})}&0&0& \ldots &\beta_{(x_{2},l_{c})}&\ldots &0&0& \ldots &\beta_{(x_{n},l_{c})}\end{array} \right]
 \end{tiny}
  \label{eqn:fsub}
 \end{equation}
 \hrule
 \end{figure*}
 where  $\b{Y}^T$ is as in (\ref{eqn:YY}). $\b{Y}$ is the vector of messages flowing through the outgoing edges of all the source nodes and is of order $((nc+\overset{n}{\underset{i=1}{\sum}} \mid K_{i} \mid)\times 1 )$. Here $K_{i,j}$ denotes the index of $j$-th message in the side information set of receiver $R_{i}$ and $\b{X} =[ x_{1} ~ x_{2} ~ x_{3}\ldots ~x_{n}]^T$ is the vector of input messages.
  The matrix $A$ is of order $ (nc+\overset{n}{\underset{i=1}{\sum}} \mid K_{i} \mid) \times n $ and it can be split in the form,
 \begin{eqnarray}
 A=\left[ \begin{array}{c}
 A_{B} \\
 A_{SI}\\
 \end{array} \right]
 \end{eqnarray}
 where $ A_{B} $  is of order $nc \times n$ and $A_{SI}$ is of order $\overset{n}{\underset{i=1}{\sum}}\mid K_{i} \mid \times n$. The matrix $A_{B}$ is a matrix formed by row-concatenation of matrices $A_{i}$, $i=1,\ldots n$ where each $A_{i}$ is a $c \times n$ matrix in which all elements in the $i$-th column are ones and the rest all are zeros as given in (\ref{ll}).
{\footnotesize
 \begin{equation}
  \begin{array}{r@{}}
   % \text{$A_{1}$}~\{ \null\\\null\\ \null\\\null\\\null\\\null\\\null\\ \null\\\null\\\null\\\null]
    \text{$A_{1}$}~\left\{\begin{array}{ccccccc}\null\\\null\\\null\\\null\\\null\end{array}\right.\\
     \text{$A_{2}$}~\left\{\begin{array}{ccccccc}\null\\\null\\\null\\\null\\\null\end{array}\right.\\
      \text{$.$}\\
      \text{$.$}\\
       \text{$A_{n}$}~\left\{\begin{array}{ccccccc}\null\\\null\\\null\\\null\\\null\end{array}\right.
  \end{array}
  \left [
   \begin{array}{cccccccccccccc}
1& 0 & 0 & 0& \ldots &0\\
1& 0 & 0 & 0& \ldots&0\\
.\\
.\\
1& 0 & 0 & 0& \ldots &0\\
0 &1& 0&0& \ldots &0 \\
0 &1& 0&0 & \ldots &0 \\
.\\
.\\
0 &1& 0&0& \ldots &0 \\
.\\
.\\
0 &0 &0 &0& \ldots &1\\
0 &0 &0 &0& \ldots &1\\
.\\
.\\
0 &0 &0 &0& \ldots &1
   \label{ll}     
    \end{array}
  \right ]
\end{equation}
}
 Each $A_{i}$ corresponds to the message passed by the source node $x_{i}$ to the intermediate nodes, $l_{j}$, $j=1,\ldots,c$. The matrix  $A_{SI}$ has only one non-zero element (which is one) in each row. This matrix corresponds to the side information possessed by the receivers and each successive set of $\mid K_{i} \mid$ rows correspond to the side information possessed by $ R_{i} $ for $i=1$ to $n$. In each set of $ \mid K_{i} \mid $ rows, each row is distinct and has only one non-zero element (which is one as we operate over the finite field $F_{2}$.) which occupies the respective column-position of one of the messages in the prior set of $R_{i}$. Hence the matrix $A$ is fixed for a fixed $c$.\\  
 \begin{figure*}
  \hrule
  \begin{equation}
  \begin{tiny}
  B_{B}=\left[ \begin{array}{ccccccccccccccccc}
\epsilon_{(l_{1},R_{1})}&0&0& \ldots &0&\epsilon_{(l_{2},R_{1})}&0&0& \ldots &0& \ldots &\epsilon_{(l_{c},R_{1})}&0&0& \ldots &0\\
0&\epsilon_{(l_{1},R_{2})}&0& \ldots &0&0&\epsilon_{(l_{2},R_{2})}&0& \ldots &0& \ldots &0&\epsilon_{(l_{c},R_{2})}&0& \ldots &0\\
 ....\\
 .....\\
 .....\\
 0&0&0& \ldots &\epsilon_{(l_{1},R_{n})}& 0&0&0& \ldots &\epsilon_{(l_{2},R_{n})}& \ldots &0&0&0& \ldots &\epsilon_{(l_{c},R_{n})}
 \end{array} \right]
 \end{tiny}
 \label{eqn:Bsub}
  \end{equation}
 \end{figure*} 
  
The matrix $F$ relates to the messages sent in the broadcast channel and the side information possessed by the the receivers and is of order $(nc+\overset{n}{\underset{i=1}{\sum}}\mid K_{i} \mid)\times  (nc+\overset{n}{\underset{i=1}{\sum}}\mid  K_{i} \mid)$. It is the matrix that satisfies the following relation.\\
  \begin{equation}
   \b{Y}' =F~\b{Y}=F~A~\b{X},
   \label{eqn:Y'}
   \end{equation}
   where $\b{Y}'^T$ is as in (\ref{eqn:YYY}). $\b{Y}'$ is the vector of messages flowing to each of the receiver. We can observe that $F$ can be split into four block matrices as given below.\\ 
  \begin{eqnarray}
                F=\left[ \begin{array}{cc}
 F_{B} & 0\\
  0  &I \end{array} \right]
  \label{eq:F}
  \end{eqnarray} 
  Matrix $F_{B}$ is a square matrix of order $nc$ which is of the form given in (\ref{eqn:fsub}) and $I$ is the identity matrix. The elements $\beta_{(x_{i},l_{j})},\forall i=1,\dots,n$ and $j=1,\ldots,c$ belong to the finite field $F_{2}$.  Every  $((i-1)n+1$)-th to $((i-1)n+n)$-th row are identical for $i=1,2,\ldots,c$. If  $((i-1)n+1$)-th row is denoted as $t_{i}$,   
  \begin{equation}
  t_{i}~A_{B}~\b{X}= g_{i}
  \end{equation}
  for $i=1,2,\ldots,c$.
  
 \begin{figure*}
   \hrule
   \vspace{.2 cm}
{\footnotesize
\begin{equation}
 F_{B}=\left[ \begin{array}{cccccccccccccccccccc}
\beta_{(x_{1},l_{1})}&0& \beta_{(x_{2},l_{1})}&0 & \beta_{(x_{3},l_{1})}&0  \\
\beta_{(x_{1},l_{1})}&0& \beta_{(x_{2},l_{1})}&0 & \beta_{(x_{3},l_{1})}&0  \\
\beta_{(x_{1},l_{1})}&0& \beta_{(x_{2},l_{1})}&0 & \beta_{(x_{3},l_{1})}&0  \\
0&\beta_{(x_{1},l_{2})}&  0&\beta_{(x_{2},l_{2})} &0&\beta_{(x_{3},l_{2})}  \\
0&\beta_{(x_{1},l_{2})}&  0&\beta_{(x_{2},l_{2})} &0&\beta_{(x_{3},l_{2})}  \\
0&\beta_{(x_{1},l_{2})}&  0&\beta_{(x_{2},l_{2})} &0&\beta_{(x_{3},l_{2})}  
\end{array} \right]  
\label{eqn:example1fsub}
\end{equation}
}
 \hrule
\end{figure*}
\begin{figure*}
{\footnotesize
\begin{equation}
  B=\left[ \begin{array}{cccccccccccccc}
\epsilon_{(l_{1},R_{1})}&0&0&\epsilon_{(l_{2},R_{1})}&0&0 & \epsilon_{(x_{2},R_{1})}&0&0\\
0&\epsilon_{(l_{1},R_{2})}&0&0&\epsilon_{(l_{2},R_{2})}&0&0& \epsilon_{(x_{3},R_{2})}&0\\
0&0&\epsilon_{(l_{1},R_{3})}&0&0&\epsilon_{(l_{2},R_{3})} &0&0& \epsilon_{(x_{1},R_{3})}
 \end{array} \right]
 \label{bexample1}
\end{equation}
}
 \hrule
\end{figure*}

 The matrix $B$ is of order $n \times (nc+\overset{n}{\underset{i=1}{\sum}}\mid K_{i} \mid)$. It relates to the decoding operations done at the receivers. It is the matrix that satisfies the following relation,\\
  \begin{equation}
   \b{Z} =B~\b{Y}'=B~F~A~\b{X},
   \label{eqn:B}
   \end{equation} 
   where $\b{Z} = [z_{1}~ z_{2} ~z_{3}\ldots~z_{n}]^T$, is the vector of output messages decoded at the receivers. The  matrix $B$ can be split into two block matrices as below.\\
 \begin{equation}
  B =\left[ \begin{array}{cc}
 B_{B} & B_{SI}  \end{array} \right] ,
 \end{equation}
where  $ B_{B}$ is a matrix of order $ n \times nc$ and in every row only $c$ elements are non-zero and the non-zero elements corresponds to whether or not $R_{i}$ uses that particular transmission to decode its wanted message. The matrix $ B_{SI}$ is of order $ n \times  \overset{n}{\underset{i=1}{\sum}}\mid K_{i} \mid$. It relates to the side information possessed by the receivers. In this matrix all elements except the $i$-th element in every successive set of $\mid  K_{i}\mid$ columns are strictly zeros, for all $i=1$ to $n$. The rest of the elements are either one or zero  and it depends on the messages used by a receiver  to decode its wanted message. The matrix $B_{B}$ is as in (\ref{eqn:Bsub}). The elements $\epsilon_{l_{j},R_{i}}$ for $j=1,\ldots, c$ and $i=1,\ldots,n$ belong to the finite field $F_{2}$. From (\ref{eqn:Y}), (\ref{eqn:Y'}) and (\ref{eqn:B}), we get\\
 \begin{equation}
   \b{Z}  =B~F~A~\b{X}.
   \label{eqn:relation}
   \end{equation}
   So,
    \begin{equation}
   M  = B~F~A.
   \label{eqn:relationm}
   \end{equation}
 An index code is solvable with $c$ number of transmissions if  we can find variables ($\beta$'s and $\epsilon$'s) such that $M$ is an identity matrix.  
%\vspace{-1 cm}

\begin{figure*}
 \hrule
\begin{eqnarray}
 \left( \begin{array}{cccccccc}
\beta_{x_{1},l_{1}} & \beta_{x_{1},l_{2}} &\ldots &\beta_{x_{1},l_{c}}\\
 \beta_{x_{2},l_{1}} & \beta_{x_{2},l_{2}}  &\ldots &\beta_{x_{2},l_{c}}\\
 .\\
 .\\
 \beta_{x_{n},l_{1}} & \beta_{x_{n},l_{2}} &\ldots &\beta_{x_{n},l_{c}}

\end{array} \right)  \left( \begin{array}{cccccc}
 \epsilon _{l_{1},R_{1}} & \epsilon _{l_{1},R_{2}} &\ldots &\epsilon _{l_{1},R_{n}} \\
 .\\
 .\\
  \epsilon _{l_{c},R_{1}} & \epsilon _{l_{c},R_{2}} &\ldots&\epsilon _{l_{c},R_{n}}
\end{array} \right)
 \label{eqn:minnum2}
\end{eqnarray}
\hrule
\end{figure*}
%%%%%%%%%?????????????????????????????????

%%%%%%%%%%%%%%%%%%%%%%%%%%%%%%%%%%%%%%%%%%%%%
\subsection{Method to Identify the Optimal Length for a Linear solution}
 \label{method}
 We have analysed the structures of the three matrices in the previous section. We need $M =  B~F~A$ to be $I$, the identity matrix. Here for a fixed length $c$, $A $ is fixed and as can be verified all the columns of $A $ are independent. Hence the rank of $A$ is $n$. So columns of $I_{n}$ (identity matrix of order $n$) lies in the column space of $A^T$. Hence the equation $A^T T_{(nc+\overset{n}{\underset{i=1}{\sum}}\mid K_{i}\mid) \times  n} =I_{n}$ has at least one solution for $T$. Observe that the number of free variables  in $T$ is  $( {n^{2}c-n^{2}+n\overset{n}{\underset{i=1}{\sum}}\mid K_{i}\mid})$ and the number of pivot variables is $n^{2}$ \cite{strang}. Hence the number of right inverses of $A^T$ is $ 2^{n^{2}c-n^{2}+n\overset{n}{\underset{i=1}{\sum}}\mid K_{i}\mid} $. We need to find a  matrix $T$ which is a right inverse of  $A^T$  as well is a product of some  $F^T $ and $B^T$ in the required form. Let us call the set of all such  matrices which satisfy both the conditions as $S(c)$. It is a function of $c$. The cardinality of the set $S(c)$ for a given length $c$ is unknown. To analyse it, let us assume that $S(c)$ is non-empty. Take a $T$ which belongs to $S(c)$. So, there exists a $B$ and $F$ such that $B ~F=  T^{T}$. Let,
\begin{equation}
 T^{T}=\left[ \begin{array}{cc}
 T_{B} & T_{SI}\end{array} \right],
 \end{equation}
 where $T_{B}$ is a $n \times nc$ matrix. Hence,
 \begin{equation}
 \left[ \begin{array}{cc}
B_{B} B_{SI} 

\end{array} \right] \left[ \begin{array}{cc}
F_{B} & 0 \\
 0  &I
\end{array} \right] =  \begin{array}{c}
 T^{T}
\end{array}  .
\end{equation} 
This gives $ T_{SI} =  B_{SI}$. So the positions which are to be strictly occupied by zeros in $ B_{SI}$ are zeros in $ T_{SI} $ also. Therefore, $ T_{SI} $ which is of order $n  \times \overset{n}{\underset{i=1}{\sum}}\mid  K_{i} \mid$  has $(n-1)(\overset{n}{\underset{i=1}{\sum}} \mid  K_{i} \mid)$ zeroes and  when the rest of the elements of $T_{SI}$ are fixed, $ B_{SI}$ also gets fixed. Keeping this in mind, we find out how many such $ T $'s are possible at the most. As the rank of  $A$  is $n$, the total number of  right inverses of  $A^{T}$  with restrictions said above (regarding the presence of zeroes at specific places) is $ 2^{n^{2}c-n^{2}+\overset{n}{\underset{i=1}{\sum}}\mid K_{i}\mid}$. Let us call this set $S'(c)$. Clearly $S(c) \subseteq   S'(c) $. Hence,\\
\begin{equation}
  \vert  S(c)  \vert   \leq  2^{n^{2}c-n^{2}+\overset{n}{\underset{i=1}{\sum}} \mid K_{i}\mid}.
  \label{23}
 \end{equation}
 We will have to identify the elements in the set $S'(c)$ which also belong to $S(c)$. But a matrix  belongs to $S(c)$ if and only if at least one pair of ($B,F$) exists such that their product is the transpose of the matrix itself. For each $T$ from $S(c)$, how many ($B,F $) pairs are possible is unknown. First of all, when we fix $T$, $ B_{SI}$ gets fixed. So for a pair ($B,F$) whose product is $T^T$ (which belongs to set $S(c)$),
  \begin{equation}
 B_{B} F_{B}=T_{B}.
 \label{FF} 
   \end{equation}          
 From (\ref{FF}) we get relations of the form,\\
 \begin{eqnarray}
 \label{eqn:columncond}
 \left[ \begin{array}{c}
 \epsilon_{l_{i},R_{1}}\\
 .\\
 .\\
 \epsilon_{l_{i},R_{n}}\end{array} \right]  
    \begin{array}{c}
 \beta_{(x_{k},l_{i})}
\end{array}   =\left[ \begin{array}{c}
 T_{col_{(k-1)c+i}}
\end{array} \right]
\end{eqnarray} 
$ \forall   k \in \lbrace1,2...n\rbrace $ and $\forall i \in \lbrace1,2....c\rbrace $
where $ T_{col_{i}}$ is the $i$-th column of $T_{B}$.
\begin{lemma}
Any matrix  $T$ which belongs to $S'(c)$ also belongs to $S(c)$  if and only if the following condition is satisfied:\\
The space spanned by the set of columns $ \lbrace   T_{col_{i}},   T_{col_{c+i}} ... T_{col_{(n-1)c+i}}  \rbrace $ in $ T_{B} $ is one or zero dimensional for all $i$.  
\end{lemma} 
 However for a $T \in S(c)$, if any such set of columns in $ T_{B} $ (i.e., the set $\lbrace T_{col_{i} } ,  T_{col_{ c+i}} , \dots, T_{col_{ (n-1)c+i}}\rbrace, \forall  i$) has only all-zero columns, then either all the $\beta $'s or $ \epsilon $'s corresponding to that set are  completely zeros. When the $\beta$'s are zeros, the $\epsilon$'s can take any of the $2^n$ values possible and vice versa. Hence the number of possibilities for such a set of all-zero columns is $2^{n+1}-1$. Hence the total number of ($B, F $) possible for a $T$ matrix is $(2^{n+1}-1)^{\lambda }$, where $\lambda $, $0 \leq \lambda \leq c$ is the number of  sets of columns whose all elements are all-zero columns among the sets $\lbrace T_{col_{i} } ,  T_{col_{ c+i}} , \dots, T_{col_{ (n-1)c+i}}  \rbrace, \forall  i $. 
\begin{theorem} A length $c$ is optimal for a linear index coding problem if and only if all the matrices in $S(c)$ have $\lambda=0$.
\end{theorem}

Theorem 1 is illustrated in Example-1 and Example-2 ~in the Appendix.

%%%%%%%%???????

%\begin{figure*}
% \hrule
%\begin{eqnarray}
% \left( \begin{array}{cccccccc}
%\beta_{x_{1},l_{1}} & \beta_{x_{1},l_{2}} &\ldots &\beta_{x_{1},l_{c}}\\
% \beta_{x_{2},l_{1}} & \beta_{x_{2},l_{2}}  &\ldots &\beta_{x_{2},l_{c}}\\
% .\\
% .\\
% \beta_{x_{n},l_{1}} & \beta_{x_{n},l_{2}} &\ldots &\beta_{x_{n},l_{c}}
%
%\end{array} \right)  \left( \begin{array}{cccccc}
% \epsilon _{l_{1},R_{1}} & \epsilon _{l_{1},R_{2}} &\ldots &\epsilon _{l_{1},R_{n}} \\
% .\\
% .\\
%  \epsilon _{l_{c},R_{1}} & \epsilon _{l_{c},R_{2}} &\ldots&\epsilon _{l_{c},R_{n}}
%\end{array} \right)
% \label{eqn:minnum2}
%\end{eqnarray}
%\hrule
%\end{figure*}
%%%%%%%%%
   
   \subsection{Minimum Number of Codes Possible for an Optimal $c$}
  In this subsection, we establish some already known results algebraically. We  find the lower bound on the number of linear codes  which are optimal in terms of bandwidth for a single unicast index coding problem and prove that this is met with equality by a special class of index coding problems. We only consider linear codes with optimal length. For the optimal $c$, the number of  matrices which are right inverses of $A^T$ and whose transpose is a product of  some $B$ and $F$ gives the number of codes possible with that length, which is also the size of  the set $S(c)$. But for any $T \in S(c)$, 
 \begin{equation}
   A_{B}^{T}   T_{B}^{T}  =  I  -  A_{SI}^{T}   T_{SI}^{T}
   \label{eqn:minnum1}
  \end{equation}\\
   where LHS will be of a form as in (\ref{eqn:minnum2}).
%%%%%%%%%????

 \begin{theorem} The number of linear index coding solutions having optimal length $c$ for a single unicast IC problem is at-least
 \begin{equation}
  {\frac{{\overset{c-1}{\underset{i=0}{\prod}}{(2^{c}-2^{i})} }}{c!}}   
  \label{eqn:minnum4}
  \end{equation}
\end{theorem}

Note that all possible matrices occupying RHS of (\ref{eqn:minnum1}) are exactly the collection of matrices which fits the index coding problem as per the definition of a fitting matrix in \cite{Birk}. Hence algebraically we have proved the already established result \cite{Birk} that the optimal length of a linear solution is the minimum among the ranks of all the matrices which fits the IC problem.\\
\begin{corollary} The number of index codes possible with the optimal length  $c$ for a single unicast IC problem is given by
 \begin{equation}
  {\frac{\mu{\overset{c-1}{\underset{i=0}{\prod}}{(2^{c}-2^{i})} }}{c!}},
  \end{equation}
  where $\mu$ is the number of  $ T_{SI}^{T}$ matrices  out of the $2^{{\overset{n}{\underset{i=1}{\sum}}} \mid  K_{i}\mid}$ possible ones which give a $c$-rank RHS  matrix of (\ref{eqn:minnum1}) with unique column space.
  \begin{proof} 
  The Proof of this follows from that of Theorem 2.
  \end{proof}
  \end{corollary}

\begin{corollary}The bound in Theorem 2 is satisfied with equality by a single unicast single uniprior problem.\\
\end{corollary}

\vspace*{-0.8cm}
%%%%%%%%%%%%%%%%%%%%%%%%%%%%%%%%%%%%%%%%%%%%%%%%%%%
% \section{Discussion}
%  For an optimal $c$, the maximum number of index codes possible is bounded by $2^{nc}$. In this paper we have given a lower bound also. This lower bound is satisfied with equality for a single unicast problem in which $\mid K_{i}\mid=1 $ and $K_{i} \cap K_{j}$=0, for $i \neq j, i,j$ = 1 to n . We would like to extend this work to find out least complexity algorithms which finds IC solutions by matrix completion. Harvey \textit{et. all} in [9] gives such algorithms for  multicast network codes. However what we have is a general problem and hence their results are not applicable. We have followed an approach which is different and simpler than Koetter and medard's [5] for this specific class of network coding problem.
%%%%%%%%%%%%%%%%%%%%%%%%%%%%%%%%%%%%%% 
% \section*{Acknowledgement}
% The authors would like to thank Anoop Thomas for the useful discussions on the topic.
%%%%%%%%%%%%%%%%%%%%%%%%%%
 
%%%%%%%%%%%%%%%%%%%%%%%%%%%%%%%%%%%%%%%%%%%
 
~~~~~

%\newpage

%%%%%%%%%%%%%%%%%%%%%%%%%%%%%%%%%%%%%%%%%%%%%%%%%%%%%%%%%%%%%%%%%%%%%%
\newpage 

~~~~~

%\newpage

%%%%%%%%%%%%%%%%%%%%%%%%%%%%%%%%%%%%%%%%%%%%%%%%%%%%%%%%%%%%%%%%%%%%%%
\begin{center}
{\bf Appendix}
\end{center}
%%%%%%%%%%%%%%%%%%%%%%%%%%%%%%%%%%%%%%%
\begin{example}
\label{sec:mn3}
Let $ m=n=3$. Each $R_{i}$ wants $x_{i}$ and knows $x_{i+1}$, where $+$ is mod-3 addition. The optimal length of a linear IC solution for this problem is $2$, which we prove in section IV. The graph G for $c=2$ is as in Fig. 2:
\end{example}

$\b{Y}^T  =[ Y((x_{1},l_{1}))~ Y((x_{1},l_{2})) ~Y((x_{2},l_{1}) ~ Y((x_{2},l_{2})\\~ Y((x_{3},l_{1}))~ Y((x_{3},l_{2})) ~ Y((x_{2},R_{1})) ~Y((x_{3},R_{2}))~ Y((x_{1},R_{3}))]$, i.e., the set of all outgoing messages from the source nodes. The vector of input messages is $\b{X} =[ x_{1}~x_{2}~x_{3}]^T$. The vector $\b{Y}'^T =[Y((l'_{1},R_{1}))~Y((l'_{1},R_{2}))~Y((l'_{1},R_{3}))Y((l'_{2},R_{1}))Y((l'_{2},R_{2}))\\~Y((l'_{2},R_{3})) ~Y((x_{2},R_{1})) ~Y((x_{3},R_{2}))~ Y((x_{1},R_{3})]$,  i.e., the vector of messages flowing to each of the receivers. The output at the receivers after decoding, is $Z=[z_{1}~ z_{2} ~z_{3}]^T$. The $A$ matrix is as  below.
 \begin{eqnarray}
  A = \left[ \begin{array}{ccc}
1 & 0 & 0\\
1& 0 & 0\\
0 & 1 & 0 \\
0 & 1 & 0 \\
0 & 0  &1 \\
0 & 0  &1 \\
0 & 1 & 0 \\
0 & 0  &1\\
1& 0 & 0
 \end{array} \right] 
\end{eqnarray}
The $F_{B}$ is as in (\ref{eqn:example1fsub}) and $B$ matrix is as in (\ref{bexample1}). The number of linear codes which are optimal in terms of length is three. They are $\mathfrak{C}_{1}:$ $ x_{1}   \oplus   x_{2} $, $ x_{2}   \oplus   x_{3} $,  $\mathfrak{C}_{2}:$ $ x_{1}   \oplus   x_{3} $, $ x_{3}   \oplus   x_{2} $,  $\mathfrak{C}_{3}:$ $ x_{1}  \oplus   x_{3} $, $ x_{1}   \oplus  x_{2} $. For the code  $\mathfrak{C}_{1}$, the matrices $F_{B}$ and $B$ are as in (\ref{eqn:example1Bsubone}). For the code  $\mathfrak{C}_{2}$, the matrices $F_{B}$ and $B$ are as in (\ref{eqn:example1Bsubone1}). For the code  $\mathfrak{C}_{3}$, the matrices $F_{B}$ and $B$ are as in (\ref{eqn:example1Bsubone2}).

\begin{figure}[htbp]
\centering
\includegraphics[scale=.75]{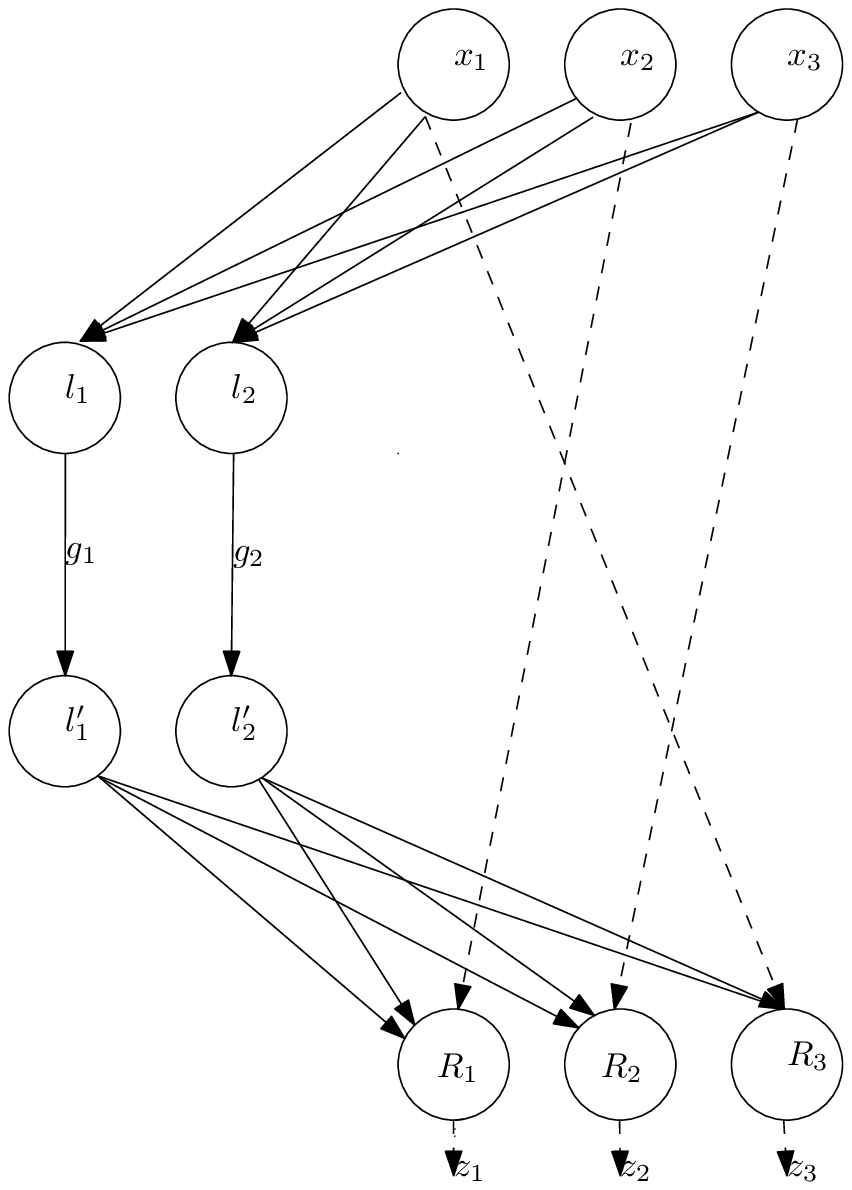}
\caption{ Equivalent network code corresponding to the IC problem in Example \ref{sec:mn3}}
\end{figure}
\begin{figure*}
\begin{eqnarray}
 F_{B}=\left( \begin{array}{cccccccccccccccccccc}
1&0& 1 &0 & 0&0  \\
1&0& 1 &0 & 0&0  \\
1&0& 1 &0 & 0&0  \\
0&0&  0&1 &0&1  \\
0&0&  0&1 &0&1  \\
0&0&  0&1 &0&1  
\end{array} \right),
& B=\left( \begin{array}{cccccccccccccc}
1&0&0&0&0&0 & 1 &0&0\\
0&0&0&0&1&0&0& 1&0\\
0&0&1&0&0&1 &0&0& 1
\end{array} \right)
\label{eqn:example1Bsubone}
\end{eqnarray}
\begin{eqnarray}
F_{B}=\left( \begin{array}{cccccccccccccccccccc}
1&0& 0 &0 & 1&0  \\
1&0& 0 &0 & 1&0  \\
1&0& 0 &0 & 1&0  \\
0&0&  0&1 &0&1  \\
0&0&  0&1 &0&1  \\
0&0&  0&1 &0&1  
\end{array} \right),
& B=\left( \begin{array}{cccccccccccccc}
1&0&0&1&0&0 & 1 &0&0\\
0&0&0&0&1&0&0& 1&0\\
0&0&1&0&0&0 &0&0& 1
\end{array} \right)
\label{eqn:example1Bsubone1}
\end{eqnarray}
\begin{eqnarray}
F_{B}=\left( \begin{array}{cccccccccccccccccccc}
1&0& 0 &0 & 1&0  \\
1&0& 0 &0 & 1&0  \\
1&0& 0 &0 & 1&0  \\
0&1&  0&1 &0&0  \\
0&1&  0&1 &0&0  \\
0&1&  0&1 &0&0  
\end{array} \right),
& B=\left( \begin{array}{cccccccccccccc}
0&0&0&1&0&0 & 1 &0&0\\
0&1&0&0&1&0&0& 1&0\\
0&0&1&0&0&0 &0&0& 1
\end{array} \right)
\label{eqn:example1Bsubone2}
\end{eqnarray}
\hrule
\end{figure*}
%%%%%%%%%%%%%%%%%%%%%%%%%%%%%%%%%%%%%%%%%%%%%
\textbf{Example 1.} \textit{(continued)}. We will illustrate Theorem 1 for the problem in Example 1. We will prove $c=1$ is not possible in this case. We can observe that $n=3$. Hence, from (\ref{23}), $ 2^3=8 $ matrices are there which belong to $S'(1)$. We found them by brute force among $2^{12}$ matrices which has zeros at places which are occupied by zeros strictly in the corresponding $B_{SI}$. Let us denote them by $ T_{1},T_{2} ,...T_{8}$. They are as given below.
 \[\left[ \begin{array}{ccc}
 1 & 0 & 0\\
 1 & 1& 0\\
 0 & 0 & 1\\
 1 & 0 & 0\\
 0 & 0& 0\\
 0 & 0& 0
  \end{array} \right],
  \left[ \begin{array}{ccc}
 1 & 0 & 1\\
 1 & 1& 0\\
 0 & 0 & 1\\
 1 & 0 & 0\\
 0 & 0& 0\\
 0 & 0& 1
  \end{array} \right],
  \left[ \begin{array}{cccccc}
 1 & 0 & 0\\
 1 & 1& 0\\
 0 & 1 & 1\\
 1 & 0 & 0\\
 0 &  1 & 0\\
 0 & 0& 0
  \end{array} \right],
  \left[ \begin{array}{cccccc}
 1 & 0 & 1\\
 1 & 1 & 0\\
 0 & 1 & 1\\
 1 & 0 & 0\\
 0 & 1& 0\\
 0 & 0& 1
  \end{array} \right]\]
  \[\left[ \begin{array}{cccccc}
 1 & 0 & 0\\
 0 & 1& 0\\
 0 & 0 & 1\\
 0 & 0 & 0\\
 0 & 0& 0\\
 0 & 0& 0
  \end{array} \right],
  \left[ \begin{array}{cccccc}
 1 & 0 & 1\\
 0 & 1 & 0\\
 0 & 0 & 1\\
 0 & 0 & 0\\
 0 & 0 & 0\\
 0 & 0& 1
  \end{array} \right],
  \left[ \begin{array}{cccccc}
 1 & 0 & 0\\
 0 & 1& 0\\
 0 & 1 & 1\\
 0 & 0 & 0\\
 0 & 1& 0\\
 0 & 0& 0
  \end{array} \right],
  \left[ \begin{array}{cccccc}
 1 & 0 & 1\\
 0 & 1 & 0\\
 0 & 1 & 1\\
 0 & 0 & 0\\
 0 & 1& 0\\
 0 & 0& 1
  \end{array} \right]\]
Denote by $T_{B,k}$, the matrix formed by taking the first $nc$ columns of $T_{k}^T$ and $T_{col_{i},k }$ is the $i$-th column of $T_{B,k}$, for $k=1,\ldots,8$. As can be seen none of the $T_{k}$ matrices satisfy the criterion of  having dimension $1$ or less for the sets of columns of $T_{B,k}$ (the set $\lbrace T_{col_{i},k },  T_{col_{ c+i},k}, \ldots, T_{col_{ {(n-1)c+i},k}} \rbrace, \forall  i$). Hence, there does not exist a solution with $c=1$.
%%%%%%%%%---------------------------------------------------------
  \begin{example}
Let $m=n=3$ and  $ R_{i} $ wants  $ x_{i} $, $\forall i \in \lbrace{1,2,3}\rbrace$. $ R_{1} $ knows $ x_{2} $ and $ x_{3} $. $ R_{2} $ knows $ x_{3} $. $ R_{3} $ knows $ x_{1} $.
 \end{example}
The optimal value of $c$ is $2$. For $c=1$, size of $S'(c)=16$ (from  (\ref{23})). The matrices $T_k,~ k =1 ,\ldots,16$ which belong to $S'(1)$ are found by brute force among $2^{13}$ matrices which has zeros at places, which are to be occupied strictly by zeros in the corresponding $B_{SI}$.  They are  :\\
  \[\left[ \begin{array}{cccccc}
 1 & 0 & 0\\
 0 & 1 & 0\\
 0 & 0 & 1\\
 0 & 0 & 0\\
 0 & 0 & 0\\
 0 & 0& 0 \\
 0& 0 & 0
  \end{array} \right],
 \left[ \begin{array}{cccccc}
 1 & 0 & 0\\
 0 & 1 & 0\\
 1 & 0 & 1\\
 0 & 0 & 0\\
 1 & 0 & 0\\
 0 & 0& 0 \\
 0& 0 & 0
  \end{array} \right],
  \left[ \begin{array}{cccccc}
 1 & 0 & 0\\
 1 & 1 & 0\\
 0 & 0 & 1\\
 1 & 0 & 0\\
 0 & 0 & 0\\
 0 & 0& 0 \\
 0& 0 & 0
  \end{array} \right],
  \left[ \begin{array}{cccccc}
 1 & 0 & 0\\
 1 & 1 & 0\\
 1 & 0 & 1\\
 1 & 0 & 0\\
 0 & 0 & 0\\
 0 & 0& 0 \\
 0& 0 & 0
  \end{array} \right]\]
 \[ \left[ \begin{array}{cccccc}
 1 & 1 & 1\\
 0 & 1 & 0\\
 0 & 1 & 1\\
 0 & 0 & 0\\
 0 & 0 & 0\\
 0 & 1 & 0 \\
 0& 0 & 1
  \end{array} \right],
  \left[ \begin{array}{cccccc}
 1 & 1 & 1\\
 1 & 1 & 0\\
 0 & 1 & 1\\
 1 & 0 & 0\\
 0 & 0 & 0\\
 0 & 1 & 0 \\
 0& 0 & 1
  \end{array} \right],
  \left[ \begin{array}{cccccc}
 1 & 1 & 1\\
 0 & 1 & 0\\
 1 & 1 & 1\\
 0 & 0 & 0\\
 1 & 0 & 0\\
 0 & 1 & 0 \\
 0& 0 & 1
  \end{array} \right],
   \left[ \begin{array}{cccccc}
 1 & 1 & 1\\
 1 & 1 & 0\\
 1 & 1 & 1\\
 1 & 0 & 0\\
 1 & 0 & 0\\
 0 & 1 & 0 \\
 0& 0 & 1
  \end{array} \right]\]
  \[\left[ \begin{array}{cccccc}
 1 & 0 & 0\\
 0 & 1 & 0\\
 0 & 1 & 1\\
 0 & 0 & 0\\
 0 & 0 & 0\\
 0 & 1 & 0 \\
 0& 0 & 0
  \end{array} \right],
  \left[ \begin{array}{cccccc}
 1 & 0 & 0\\
 1 & 1 & 0\\
 0 & 1 & 1\\
 1 & 0 & 0\\
 0 & 0 & 0\\
 0 & 1 & 0 \\
 0& 0 & 0
  \end{array} \right],
  \left[ \begin{array}{cccccc}
 1 & 0 & 0\\
 0 & 1 & 0\\
 1 & 1 & 1\\
 0 & 0 & 0\\
 1 & 0 & 0\\
 0 & 1 & 0 \\
 0& 0 & 0
  \end{array} \right],
  \left[ \begin{array}{cccccc}
 1 & 0 & 0\\
 1 & 1 & 0\\
 1 & 1 & 1\\
 1 & 0 & 0\\
 1 & 0 & 0\\
 0 & 1 & 0 \\
 0& 0 & 0
  \end{array} \right]\]
  \[\left[ \begin{array}{cccccc}
 1 & 0 & 1 \\
 0 & 1 & 0\\
 0 & 0 & 1\\
 0 & 0 & 0\\
 0 & 0 & 0\\
 0 & 0& 0 \\
 0& 0 & 1
  \end{array} \right],
   \left[ \begin{array}{cccccc}
 1 & 0 & 1 \\
 1 & 1 & 0\\
 0 & 0 & 1\\
 1 & 0 & 0\\
 0 & 0 & 0\\
 0 & 0& 0 \\
 0& 0 & 1
  \end{array} \right],
   \left[ \begin{array}{cccccc}
 1 & 0 & 1 \\
 0 & 1 & 0\\
 1 & 0 & 1\\
 0 & 0 & 0\\
 1 & 0 & 0\\
 0 & 0& 0 \\
 0& 0 & 1
  \end{array} \right],
   \left[ \begin{array}{cccccc}
 1 & 0 & 1 \\
 1 & 1 & 0\\
 1 & 0 & 1\\
 1 & 0 & 0\\
 1 & 0 & 0\\
 0 & 0& 0 \\
 0& 0 & 1
  \end{array} \right]\]

  As can be seen none of the $T_{k}$ matrices satisfy the criterion of  having dimension $1$ for the sets of columns of $T_{B,k}$ (the set $\lbrace T_{col_{i},k } ,  T_{col_{ c+i},k} , \dots, T_{col_{ (n-1)c+i},k} \rbrace, \forall  i$). Hence $c =1$ is not a feasible length for this case. If $c=3$ is taken, one would get a matrix $T$ which belongs to the set $S(3)$, as in (\ref{ls}). For this matrix, $\lambda \neq0$. Also dimension of every set of columns (i.e., the set $\lbrace T_{col_{i} } ,  T_{col_{ c+i}} , \dots, T_{col_{ (n-1)c+i}}  \rbrace,~\forall  i$) is $1 $ or $0$. Hence $c=3$ is not optimal. Therefore, $c=2 $ should be the optimal length.\\
  \vspace{1 cm}
  \begin{eqnarray}
   \left[ \begin{array}{cccccc}
 1 & 0 & 1 \\
 0 & 0 & 0\\
 0 & 0 & 0\\
 1 & 0 & 1\\
 0 & 1 & 1\\
 0 & 0& 0 \\
 0& 0 & 0\\
 0 &1 &1\\
 0 & 0&0 \\
 1 & 0 & 0\\
 0 & 1 & 0\\
 0 & 0& 0\\
 0 & 0 & 1
  \end{array} \right]
  \label{ls}
  \end{eqnarray}

%%%----------------------------
 \begin{example}
 Let $m=n=4$. $R_{i}$ wants $x_{i}$ and knows $x_{i+1}$ where $+$ is modulo-4 operation. $x_{3}$ knows $x_{1}$ also.
 \label{mu}
 \end{example}
 The optimal length is $c=3$ and it can be checked that  $\mu=2$. The number of optimal linear codes are 56 in number thus satisfying corollary 2.
%%%%%%%%%%%%%%%%%%%%%%%%%%%%%%%%%%%%%

Example 1. was a single unicast single uniprior problem. The optimal length is $c=2$ and three solutions are possible with that length, satisfying Corollary 2.
  \begin{example}
 Let $m=n=4$. $R_{i}$ wants $x_{i}$ and knows $x_{i+1}$, where $+$ is modulo-4 addition.
  \label{eg:unip}
  \end{example}
  Here all possible matrices of the form (\ref{eqn:minnum3}) denoted by $L_{i}$, $i=1,\ldots,16$ are as in Table \ref{table1}.
  \begin{table*}
\centering
\tiny
\begin{tabular}{ccccc}
$L_{1} = \left[\begin{array}{ccccccc}
1 & 0 & 0 & 0 \\
 0& 1 &0 &0\\
 0& 0&1&0\\
 0&0&0&1
\end{array}\right]$, & $L_{2} = \left[\begin{array}{ccccccc}
 1 & 0 & 0 & 1 \\
 0& 1 &0 &0\\
 0& 0&1&0\\
 0&0&0&1
\end{array}\right]$, &
$L_{3} = \left[\begin{array}{ccccccccc}
 1 & 0 & 0 & 1 \\
 1& 1 &0 &0\\
 0& 0&1&0\\
 0&0&0&1
\end{array}\right]$,  &
$L_{4} = \left[\begin{array}{ccccccccc}
  1 & 0 & 0 & 1 \\
 1& 1 &0 &0\\
 0& 1&1&0\\
 0&0&0&1
\end{array}\right]$, \\
$L_{5} = \left[\begin{array}{ccccccccc}
  1 & 0 & 0 & 1 \\
 1& 1 &0 &0\\
 0& 1&1&0\\
 0&0&1&1
\end{array}\right]$,  &
$L_{6} = \left[\begin{array}{ccccccccc}
  1 & 0 & 0 & 1 \\
 0& 1 &0 &0\\
 0& 1&1&0\\
 0&0&0&1
 \end{array}\right]$,  &
$L_{7} = \left[\begin{array}{ccccccccc}
  1 & 0 & 0 & 1 \\
 0& 1 &0 &0\\
 0& 1&1&0\\
 0&0&1&1
 \end{array}\right]$,  &
$L_{8} = \left[\begin{array}{ccccccccc}
  1 & 0 & 0 & 1 \\
 0& 1 &0 &0\\
 0& 0&1&0\\
 0&0&1&1
 \end{array}\right]$,  \\
$L_{9} = \left[\begin{array}{ccccccccc}
  1 & 0 & 0 & 0 \\
 1& 1 &0 &0\\
 0& 0&1&0\\
 0&0&0&1
 \end{array}\right]$,  &
$L_{10} = \left[\begin{array}{ccccccccc}
  1 & 0 & 0 & 0 \\
 1& 1 &0 &0\\
 0& 1&1&0\\
 0&0&0&1
 \end{array}\right]$, &
$L_{11} = \left[\begin{array}{ccccccccc}
  1 & 0 & 0 & 0 \\
 1& 1 &0 &0\\
 0& 1&1&0\\
 0&0&1&1
 \end{array}\right]$,  &
$L_{12} = \left[\begin{array}{ccccccccc}
  1 & 0 & 0 & 0 \\
 1& 1 &0 &0\\
 0& 0&1&0\\
 0&0&1&1
 \end{array}\right]$, \\
$L_{13} = \left[\begin{array}{ccccccccc}
  1 & 0 & 0 & 0 \\
 0& 1 &0 &0\\
 0& 1&1&0\\
 0&0&0&1
 \end{array}\right]$,  &
$L_{14} = \left[\begin{array}{ccccccccc}
  1 & 0 & 0 & 0 \\
 0& 1 &0 &0\\
 0& 1&1&0\\
 0&0&1&1
 \end{array}\right]$, &
$L_{15} = \left[\begin{array}{ccccccccc}
  1 & 0 & 0 & 0 \\
 0& 1 &0 &0\\
 0& 0&1&0\\
 0&0&1&1
 \end{array}\right]$,  &
$L_{16} = \left[\begin{array}{ccccccccc}
  1 & 0 & 0 & 1 \\
 1& 1 &0 &0\\
 0& 0&1&0\\
 0&0&1&1
 \end{array}\right]$  
 \end{tabular}
\caption{\small  Fitting matrices for Example \ref{eg:unip}}
\label{table1}
\hrule
 \end{table*}
  Only  $L_{5}$ has dimension four. The set of all optimal index codes is given by the collection of all possible basis of the column space of this matrix. They are $28$ in number. Hence corollary 2 is satisfied. We list out those codes in Table \ref{table2}.

  \begin{table*}
\label{Tab:4userTable}
\scriptsize
\centering{}
\begin{tabular}{|c|c|}
\hline
 {Code} &  {Encoding} \\
 \hline
$\mathfrak{C}_{1}$ & $x_{1}+x_{2},x_{2}+x_{3},x_{3}+x_{4}$   \\
\hline
$\mathfrak{C}_{2}$ & $x_{1}+x_{2},x_{2}+x_{3},x_2+x_4$   \\
\hline
$\mathfrak{C}_{3}$ & $x_{1}+x_{2},x_{2}+x_{3},x_1+x_2+x_3+x_4$  \\
\hline
$\mathfrak{C}_{4}$ & $x_{1}+x_{2},x_{2}+x_{3},x_1+x_4$   \\
\hline
$\mathfrak{C}_{5}$ & $x_{1}+x_{2},x_{3}+x_{4},x_1+x_3$   \\
\hline
$\mathfrak{C}_{6}$ & $x_{1}+x_{2},x_{3}+x_{4},x_2+x_4$   \\
\hline
$\mathfrak{C}_{7}$ & $x_{1}+x_{2},x_{3}+x_{4},x_1+x_4$   \\
\hline
$\mathfrak{C}_{8}$  & $x_{1}+x_{2},x_{1}+x_{3},x_2+x_4$  \\
\hline
$\mathfrak{C}_{9}$ & $x_{1}+x_{2}, x_{1}+x_{3},x_1+x_2+x_3+x_4$   \\
\hline
$\mathfrak{C}_{10}$ & $x_{1}+x_{2}, x_{1}+x_{3},x_1+x_4$  \\
\hline
$\mathfrak{C}_{11}$ & $x_{1}+x_{2},x_{2}+x_{4},x_1+x_2+x_3+x_4$   \\
\hline
$\mathfrak{C}_{12}$ & $x_{1}+x_{2},x_1+x_2+x_3+x_4,x_1+x_4$   \\
\hline
$\mathfrak{C}_{13}$  & $x_{2}+x_{3},x_3+x_4,x_1+x_3$  \\
\hline
$\mathfrak{C}_{14}$ & $x_{2}+x_{3},x_3+x_4,x_1+x_2+x_3+x_4$  \\
\hline
$\mathfrak{C}_{15}$ & $x_{2}+x_{3},x_3+x_4,x_1+x_4$  \\
\hline
$\mathfrak{C}_{16}$ & $x_{2}+x_{3},x_1+x_3,x_2+x_4$  \\
\hline
$\mathfrak{C}_{18}$ & $x_{2}+x_{3},x_1+x_3,x_1+x_4$   \\
\hline
$\mathfrak{C}_{19}$  & $x_{2}+x_{3},x_2+x_4,x_1+x_2+x_3+x_4$   \\
\hline
$\mathfrak{C}_{20}$  & $x_{2}+x_{3},x_2+x_4,x_1+x_4$  \\
\hline
$\mathfrak{C}_{21}$ & $x_{3}+x_{4},x_1+x_3,x_2+x_4$   \\
\hline
$\mathfrak{C}_{22}$  & $x_{3}+x_{4},x_1+x_3,x_1+x_2+x_3+x_4$  \\
\hline
$\mathfrak{C}_{23}$ & $x_{1}+x_{3},x_2+x_4,x_1+x_4$  \\
\hline
$\mathfrak{C}_{24}$ & $x_{1}+x_{3},x_1+x_2+x_3+x_4,x_1+x_4$   \\
\hline
$\mathfrak{C}_{25}$ & $x_{2}+x_{4},x_1+x_2+x_3+x_4,x_1+x_4$   \\
\hline
$\mathfrak{C}_{26}$ & $x_{3}+x_{4},x_2+x_4,x_1+x_2+x_3+x_4$  \\
\hline
$\mathfrak{C}_{27}$  & $x_{3}+x_{4},x_2+x_4,x_1+x_4$   \\
\hline
$\mathfrak{C}_{28}$  & $x_{3}+x_{4},x_1+x_2+x_3+x_4,x_1+x_4$   \\
\hline
\end{tabular}
\caption{\small{ All possible optimal linear solutions for Example \ref{eg:unip}}.}
\hrule
\label{table2}
\end{table*}

%%%%%%%%%%????????????

%%%%%%%%%%%%%%%%%%%%%%%%%%%%%%%%%%%%%%%%%
\section{Proof of Lemma 1}

 Proof of only-if part: If $T\in S(c)$, From (\ref{eqn:columncond}), we get relations of the form as below.
 \begin{equation}
 \left[ \begin{array}{c}
 \epsilon_{l_{i},R_{1}}\\
 \epsilon_{l_{i},R_{2}}\\
 .\\
 .\\
 \epsilon_{l_{i},R_{n}}

\end{array} \right]  \begin{array}{cc}
 \beta_{(x_{k},l_{i})}
\end{array}   =\left[ \begin{array}{c}
 T_{col_{(k-1)c+i}}.\\
\end{array} \right]
\end{equation}
Also,
\begin{equation}
 \left[ \begin{array}{c}
 \epsilon_{l_{i},R_{1}}\\
 \epsilon_{l_{i},R_{2}}\\
 .\\
 .\\
 \epsilon_{l_{i},R_{n}}

\end{array} \right]   \begin{array}{cc}
 \beta_{(x_{k'},l_{i})}
\end{array}   =\left[\begin{array}{c}
 T_{col_{(k'-1)c+i}}.
\end{array} \right]
\end{equation}
\\
Hence $  T_{col_{(k'-1)c+i}} $ has to be expressible as a multiple of $ T_{col_{(k-1)c+i}}$ or vice verse, $ \forall $ $ k $, $ k'\in \lbrace1,2...n\rbrace $ and for every $i\in\lbrace 1,2...c\rbrace$. This is not possible unless any such set of columns is one dimensional or has only all-zero columns which makes it zero dimensional. \\
Proof of if part : If the space spanned by the set of columns $ \lbrace   T_{col_{i}},   T_{col_{c+i}} ... T_{col_{(n-1)c+i}}  \rbrace $ in $ T_{B} $ is one or zero dimensional for all $i$ for a $T \in S'(c)$, one can  always find values for variables ($\epsilon$'s and $\beta$'s) satisfying (\ref{eqn:columncond}) for each of these sets. Hence one can get a pair $(B,F)$ such that (\ref{FF}) is satisfied by substituting these values. Hence $T\in S(c)$. Hence the proof is complete.

 \begin{figure*}
 \vspace{-.5 cm}
  \begin{eqnarray}
\left[ \begin{array}{cccccccccc}
1 & 0 &0 &\ldots &p_{ {1, \lbrace j :K_{j}=x_{1}\rbrace}} & 0 &\ldots &0\\
 0  &  1 & 0& \ldots &p_{{2,\{ j' :K_{j'}=x_{2} \}}} & 0& \ldots &0\\
 .\\
 .\\
 0 &0 &0 &\ldots &p_{{n,\{ j'' :K_{j''}=x_{n}\}}} & 0&\ldots &1
\end{array} \right]
 \label{eqn:minnum3}
\end{eqnarray}
\hrule
\end{figure*}

\section{Proof of Theorem 1}

\begin{proof}
proof for only if part: We need to prove that if there exists a $T \in S(c) $ whose $\lambda \neq 0$ for a particular length $c$, then $c$ is not the optimal transmission length. When such a set exists, as described above, either all the $\beta$'s or $ \epsilon $'s corresponding to that are  completely zeros. If all the $ \epsilon $ are zeroes, that means that  one particular transmission is not even used by any of the receivers. Else if all the $\beta$'s corresponding are kept zeroes, then we transmit no message in one particular transmission.  So we can remove at least one transmission. Hence the proof of only if part is complete.\\
  The proof for if part goes as follows: We prove this by contradiction. Assume that a length $c$ exists such that it is feasible but not optimal and all the matrices in $S(c)$ have $\lambda=0$. Assume further that $c'= c-r $ for some $r > 0$, is  the optimal length. Then take one feasible solution with length $c'$. Add  extra $nr$ rows to the corresponding $F_{B}$ matrix and some extra $nc$ all zero columns to $B_{B}$. Let us call the new matrices $F'_{B}$ and $B'_{B}$. Let $g'_{i}, i=1,\ldots c$ be the set of broadcast messages given by $F'_{B}$ and $g_{i}$ be those which are given by $F_{B}$. One can observe that $\lbrace g'_{1},g'_{2},\ldots, g'_{c}\rbrace$ is nothing but $\lbrace g_{1},g_{2},\ldots,g_{c'}\rbrace$ plus some additional information. Hence when one sends  $\lbrace g'_{1},g'_{2},\ldots, g'_{c}\rbrace$, the receivers get whatever they would have got if $\lbrace g_{1},g_{2},\ldots,g_{c'}\rbrace$ was sent. Hence even if they do not use the extra transmissions given by $F'_{B}$, they will be able to decode their wanted messages. Hence the product of $F'_{B}$ and $B'_{B}$ matrices should belong to $S(c)$ (as it is a feasible index code) and has $\lambda \neq 0$, which is a contradiction.  Hence $c$ is the optimal length.\\
\end{proof}

\section{Proof Theorem 2}

\begin{proof}:
 Consider (\ref{eqn:minnum1}) and (\ref{eqn:minnum2}). Here if both RHS of (\ref{eqn:minnum1})  and first matrix in (\ref{eqn:minnum2}) are fixed, solution which is the second matrix in (\ref{eqn:minnum2})  will exist only if the column space of RHS of (\ref{eqn:minnum1}) is spanned by the columns of first matrix in  (\ref{eqn:minnum2}). But the rank of the first matrix in (\ref{eqn:minnum2}) is atmost $c$. Hence this is possible only if the rank of the RHS matrix in  (\ref{eqn:minnum1}) is less than or equal to $c$. The number of possible  $ T_{SI}^{T}$ matrices is $2^{{\overset{n}{\underset{i=1}{\sum}}} \mid  K_{i}\mid}$. As we know $c$ is the optimal length, there should be at least one $ T_{SI}^{T}$ such that RHS of (\ref{eqn:minnum1}) is of rank $c$. For any such RHS of (\ref{eqn:minnum1}), we can take the first matrix in (\ref{eqn:minnum2}) in $(2^{c}-1){\overset{c-1}{\underset{i=1}{\prod}}} (2^{c}-1 -\dbinom{i}{i}-\dbinom{i}{i-1}.......-\dbinom{i}{1})$ ways such that the column spaces of both the matrices are same. Each such matrix is an index code, which is feasible, and each column of the  matrix represents a transmission. As order of transmission does not matter, we need to neglect those matrices which are column-permuted versions of one another. Hence, total number of distinct transmission schemes possible is  ${\frac{(2^{c}-1)} {c!}}{\overset{c-1}{\underset{i=1}{\prod}}} (2^{c}-1 -\dbinom{i}{i}-\dbinom{i}{i-1}.......-\dbinom{i}{1})= {\frac{{\overset{c-1}{\underset{i=0}{\prod}}{(2^{c}-2^{i})} }}{c!}}   $. But there may be more than one $T_{SI}^{T}$ matrices which are of rank $c$ and whose column spaces are different. Hence the total number of index codes possible can be more than (\ref{eqn:minnum4}) also as we take into account all possible basis sets of each of the different column spaces. Example 3 is such a case. Hence (\ref{eqn:minnum4}) is a lower bound on the number of index codes possible.
\end{proof}

\section{Proof of Corollary 2}

\begin{proof}: For a single unicast single uniprior problem the RHS of (\ref{eqn:minnum1}) will be of the form (\ref{eqn:minnum3}), where all $p_{i,\{ j: K_{j}=x_{i}\}}$ for $i=1,\ldots, n$ can be $1$ or $0$. Hence total number of matrices that can be of the form (\ref{eqn:minnum3}) is $ 2^{n}$.

As can be verified only one matrix among them has rank equal to $n-1$, which is the optimal transmission length for this single unicast problem and that one matrix is that whose all $p_{i,\{j:K_{j}=x_{i}\}}$ values are one. We will prove this by contradiction. Suppose any other matrix exists with atleast one $x_{i,j}$ zero and is of rank $n-1$, it means that receiver $R_{j}$ does not use its side information $x_{i}$. This is equivalent to the case where $R_{j}$ does not have any prior information. For this case, the optimal length of transmission is $n$, which is a contradiction. Hence the number of optimal index codes is exactly what is given by (\ref{eqn:minnum4}).
\end{proof}

%%%>>>>>>>>>>>>>>>>>>

\end{document}